# Impact of Short-Range Scattering on the Metallic Transport of Strongly Correlated 2D Holes in GaAs Quantum Wells

Nicholas J. Goble,[1] John D. Watson,[2,3] Michael J. Manfra,[2,3,4,5] and Xuan P. A. Gao[1,*]

[1]*Department of Physics, Case Western Reserve University, Cleveland, Ohio 44106, USA*
[2]*Department of Physics, Purdue University, West Lafayette, Indiana 47907, USA*
[3]*Birck Nanotechnology Center, Purdue University, West Lafayette, Indiana 47907, USA*
[4]*School of Materials Engineering, Purdue University, West Lafayette, Indiana 47907, USA*
[5]*School of Electrical and Computer Engineering, Purdue University, West Lafayette, Indiana 47907, USA*

Understanding the non-monotonic behavior in the temperature dependent resistance, $R(T)$, of strongly correlated two-dimensional (2D) carriers in clean semiconductors has been a central issue in the studies of 2D metallic states and metal-insulator-transitions. We have studied the transport of high mobility 2D holes in 20nm wide GaAs quantum wells (QWs) with varying short-range disorder strength by changing the Al fraction $x$ in the $Al_xGa_{1-x}As$ barrier. Via varying the short range interface roughness and alloy scattering, it is observed that increasing $x$ suppresses both the strength and characteristic temperature scale of the 2D metallicity, pointing to the distinct role of short-range versus long-range disorder in the 2D metallic transport in this correlated 2D hole system with interaction parameter $r_s \sim 20$.

For the past thirty years, two-dimensional (2D) quantum systems have been a rich area of concentrated study for both theorists and experimentalists to explore the interplay between Coulomb interaction and disorder effects [1-4]. In the case of 2D electrons with ultra-low density and weak disorder, the quantum and strongly interacting nature of the systems becomes so prominent that various complex quantum phases and phase transitions may exist according to theory [5-11]. Obtaining a clear understanding of such strongly correlated 2D systems thus remains to be extremely important in the field of many-body physics.

In 2D electron or hole samples with high mobility, an intriguing metal to insulator transition (MIT) was observed in zero magnetic field ($B=0$) with the carrier density as the tuning parameter [2, 3, 12]. Although strong electron-electron interactions are believed to be an essential factor in the origin of this MIT in 2D, the effects of disorder seem to be non-negligible and must be incorporated in order to reconcile the subtle differences in all the 2D MIT experiments over a range of disorder and interaction strength [2, 3, 13-15]. While the understanding of the 2D MIT in the critical regime continues to advance [13-17], the mechanism of the 2D metallic conduction in the metallic regime (resistivity $\rho \ll h/e^2$) remains an outstanding problem under debate [4, 18]. Since metallic transport is the central phenomenon that challenges conventional wisdom based on localization and weakly interacting Fermi liquid theory, its understanding would shed light on the 2D MIT and transport of correlated 2D electron fluids in general. After extensive transport studies over the past two-decades, a salient feature of 2D metallic transport has now emerged: when measured over a broad temperature range from $T \ll T_F$ to $T \sim T_F$, where $T_F$ is the Fermi temperature, a non-monotonic behavior is commonly found in the temperature dependent resistivity [19]. Specifically, as phonon scattering is reduced at low temperatures, electron-impurity scattering and electron-electron interactions contribute more to the longitudinal resistance $R_{xx}$. The electronic contribution to the resistance behaves non-monotonically, first increasing and then decreasing as $T$ is lowered past a characteristic temperature $T_0$, which is comparable to $T_F$. This feature, better observed in low density 2D systems, occurs when carriers become semi-degenerate and has been observed in the three most commonly studied 2D systems: p-GaAs [19-22], n-Si [23, 24] and n-GaAs [25, 26]. The mechanism of this non-monotonic $R_{xx}(T)$ has been a key point in several leading theories of the 2D metallic transport [7,8,27-29] and is the subject of this experimental study.

In this work, we investigate how the non-monotonic temperature dependent transport associated with the metallic transport in a 2D hole system (2DHS) is impacted by the short-range disorder scattering in a series of modulation doped GaAs/$Al_xGa_{1-x}$As QWs with varying barrier height. In these quantum wells, it is well known that a change in the aluminum concentration adjusts the short-range disorder potential via two important scattering mechanisms: interface roughness and alloy scattering [30], thus providing an opportunity to systematically investigate the effect of controlled short range disorder strength on the 2D metallic behavior in the same QW or hetero-interface system. We find that tuning the strength of the short-range disorder potential has a marked impact on both the strength and temperature scale of the non-monotonic $R_{xx}(T)$. Nevertheless, the overall shape of the $R_{xx}(T)$ does not change in a qualitative manner. These findings demonstrate that the nature of disorder (e.g. short-range vs. long-range scattering) is an important ingredient in a quantitative understanding of the 2D metallic conduction with resistivity $\rho \ll h/e^2$, as pointed out in a previous report [31] where different types of 2D semiconductor hetero-interfaces were analyzed. .

Our experiments were performed on high mobility, low-density 2DHS's in 20nm wide GaAs asymmetrically doped quantum wells located 190nm below the surface [32]. The samples were grown in the (001) direction using molecular beam epitaxy (MBE). The short-range disorder potential was varied by controlling the Al mole fraction, $x$, in the $Al_xGa_{1-x}As$ barrier. Measured $x$'s were 0.07 (7%), 0.10 (10%), and 0.13 (13%) with δ-doping setback distances, $d$, of 80, 110, and 110 nm respectively. The ungated samples have hole mobilities, $\mu \approx 0.5 - 1 \times 10^6 cm^2/Vs$.

Square samples were prepared with InZn contacts annealed in a rapid thermal annealer at 450°C for 7 min in $H_2/N_2$ forming gas. The contacts were arranged in a Van der Pauw geometry, with current applied through two contacts on one side and voltage measured over the opposite side, as shown in the schematic inset in Fig.1. The sample had an approximate total area of 0.1 cm$^2$. The hole density $p$ was lowered by applying a backgate voltage with which we were able to obtain a range of $p$ from $2.23 \times 10^{10}$ cm$^{-2}$ to $1.09 \times 10^{10}$ cm$^{-2}$ (corresponding ratio between inter-hole Coulomb repulsion energy and Fermi energy $r_s$=17–25 using effective hole mass $m^*$=0.3m$_e$). The samples were measured using lock-in techniques with an excitation current on the order of 1 nA at 7Hz. With such small heating power (~10$^{-14}$Watt/cm$^2$) applied to the sample, we were able to cool the sample to temperatures as low as 50mK in a $^3$He/$^4$He dilution refrigerator without overheating the 2D holes by more than a few mK [33].

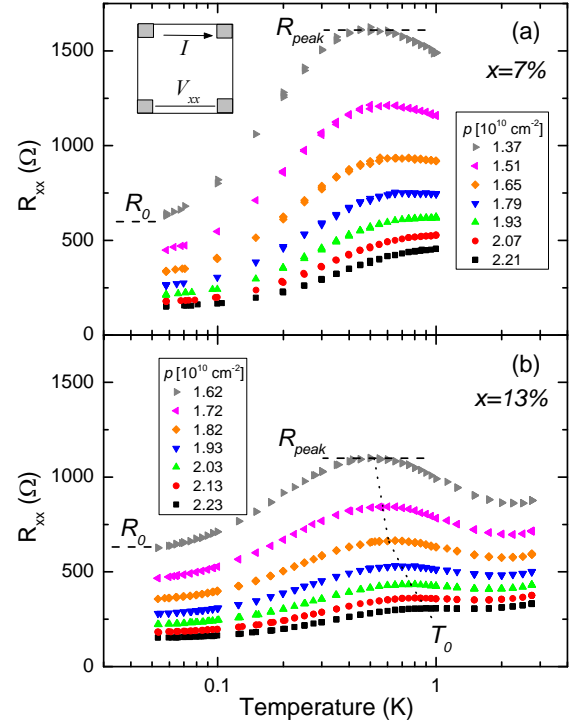

FIG. 1. Longitudinal resistance $R_{xx}$ vs. $T$ of 2D holes in a 20 nm wide GaAs quantum well with Al mole fractions of (a) 7% and (b) 13% in the $Al_xGa_{1-x}As$ barrier. The dotted lines mark $T_0$, the position of the peak in $R_{xx}(T)$.

Figure 1 shows the raw $R_{xx}(T)$ data for a sample with 7% (a) and 13% (b) Al mole fractions in the barriers. Longitudinal resistance was measured from low temperatures, $T$~50 mK, up to high temperatures, $T$=1-4 K. Varying the carrier density predictably changes the non-monotonic behavior by proportionally shifting the characteristic temperature $T_0$, at which $dR_{xx}/dT=0$. This defines the position of the non-monotonic peak of interest. Similar to previous experiments [21], $T_0$ becomes higher when the hole density increases. It has been shown that the temperature dependence of $R_{xx}$ on the metallic side of the MIT follows [19]

$$R_{xx}(T) = R_0 + R_{ph}t^3/(1+t^2) + R_{el}(T), \quad (1)$$

where $R_0$ is the residual resistance as $T \to 0$, $R_{ph}$ is the Bloch-Gruneisen resistance, $t \equiv T/T_{ph}$, $T_{ph}$ is the Bloch-Gruneisen temperature, and $R_{el}$ is the $T$-dependent part of the electronic contribution to $R_{xx}$, which can be fit to the empirical form [19]

$$R_{el}(T) = R_a x^{-1}(\alpha + x^{-2})^{-\frac{1}{2}}\exp(-x^{-1}), \quad (2)$$

where $x = kT/E_a$, $E_a$ is the activation energy and is proportional to $T_0$, and $\alpha$ is a constant. As seen in Fig.1b, at temperatures higher than 2-4K, $R_{xx}(T)$ increases with $T$. This is due to the phonon scattering term dominating over the electronic term above the Block-Gruneisen temperature [19]. At lower temperature where the phonon term diminishes rapidly (according to $\sim T^3$ power law dependence), the electronic term $R_0 + R_{el}(T)$ shows a non-monotonic behavior: $R_{xx}$ first rises to $R_{peak}=R_{xx}(T_0)$, the maximal value of $R_{xx}$ at $T_0$, then drops towards $R_0$ as $T$ reduces towards zero.

Comparing the $R_{xx}(T)$ curves in Fig.1a and b suggests that the non-monotonic electronic impurity scattering peak is more pronounced in the sample with 7% Al in the barrier. To further elaborate and quantify this effect, we systematically studied the effect of short-range scattering on the non-monotonic $R_{xx}(T)$ peak for three different Al percentages. Figure 2a shows the temperature dependent resistance normalized over $R_0$ for samples with 7%, 10% and 13% Al in the barrier, respectively. Note that to clearly and reliably illustrate the evolution of $R_{xx}(T)$ peak against Al%, curves for similar hole density (1.71, 1.73 and 1.77×$10^{10}$ cm$^{-2}$) are presented, and the curve for each Al mole fraction was obtained by averaging over different current-voltage contact arrangements as per the Van der Pauw method and two different samples. $R_0$ was obtained by fitting the curve to Eq.1 and hole density was obtained by the positions of Shubnikov de Haas oscillations at low $T$. Figure 2a shows that the relative strength of the non-monotonic $R_{xx}(T)$ peak does become stronger at lower Al%, although it is harder to identify it in the raw $R_{xx}(T)$ data (Fig.1a) due to differences in the absolute values of the resistance (or mobility). Note that, to reduce data manipulation, we did not subtract a phonon scattering term from the raw resistance data in Fig.1a. However, separating the electronic scattering from phonon scattering in $R_{xx}(T)$ does not change the curves in any significant way [34]. Briefly, fitting to Eq.1 was first conducted to determine how well the data fit the model described in Ref. [19]. As shown by the solid lines in Fig.2a, the fits agree well with the measured behavior. Using methods discussed by Mills *et al.* in Ref. [19], we isolated the electronic scattering part, which does not differ much from Fig.1a, as shown in Ref. [34]. In addition, fitted parameters ($R_0$, $R_{ph}$, $R_a$, $T_{ph}$, and $E_a$) and constant ($\alpha = 2.5$ as suggested by Ref. [19]) from Eq.1 and Eq.2 were extremely consistent with those in Ref. [19]. We also find that $\alpha$ does not vary significantly between samples with different aluminum concentrations and in an effort to reduce fitting parameters, choose to keep it constant. The difference between hand-selecting and fitting $R_0$ and $R_a$ was insignificant.

Such analysis was conducted across our range of hole densities, and $R_{peak}/R_0$ values were extracted after removing the phonon scattering term and plotted for various different hole densities for the three sets of quantum wells with different Al fractions mentioned above. The results are shown in Fig. 2b. Averaging over different current-voltage contact configurations and similar hole densities is taken into account as the error bars. Due to the imperfect contact alignment and perhaps slight inhomogeneity in carrier density and mobility in the wafer, the metallicity strength parameter, $R_{peak}/R_0$, varies somewhat between different samples with the same Al% or current-voltage contact arrangements on the same square Van der Pauw sample. However, it is still clear that $R_{peak}/R_0$ increases from ~1.7 in QWs with 13% Al in barrier to ~2.7 in QWs with 7% Al in barrier. Therefore, Fig. 2b demonstrates that the behavior described in Fig.1 and Fig. 2a is true for our entire range of densities: increased interface roughness in samples with higher Al fraction in the barrier suppresses 2D metallicity behavior for high mobility, low density 2DHS in GaAs QWs.

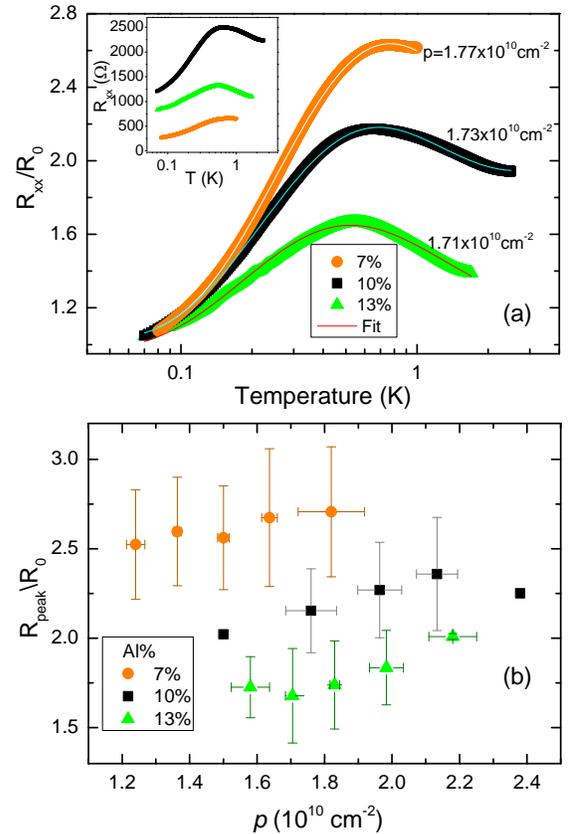

FIG. 2. (a) The temperature dependent resistance, normalized with $R_0$, for 20nm wide GaAs quantum wells with 7%, 10%, and 13% Al mole fractions at $p = 1.77$, 1.73, and $1.71 \times 10^{10}$ cm$^{-2}$, respectively. The model fit of Eq. (1) is indicated by the solid lines. The inset plots the un-normalized resistance as a function of temperature. (b) Dependence of the strength of 2D metallicity as defined by the ratio of resistance at the peak of $R(T)$ and residual resistance on the hole density. The error bar comes from averaging measurements from different current/voltage configuration and different samples. Zero-bias densities are the maximum densities for each Al%.

In addition to the strength of the non-monotonic resistance peak, we studied the position of the peak, depicted by the characteristic temperature, $T_0$, defined as the temperature where $dR_{xx}/dT = 0$. From Fig. 1, it can be seen that $R_{xx}$ changes its behavior from metallic to insulating-like above $T_0$. The characteristic temperature becomes larger when $p$ increases, which is consistent with previous findings in the literature [19,21]. Despite the obvious importance of $T_0$ in any theory about the 2D metallic conduction and MIT, it is surprising to us that there is no systematic experimental investigation on what other parameters besides the carrier density (e.g. interaction strength, mobility) control $T_0$. Figure 3 shows $T_0$ as a function of hole density for all the $R_{xx}(T)$ measurements. Different markers indicate different samples or different Van der Pauw measurement configurations for the same sample, and shaded areas highlight trends. Similar to the shift in $R_{peak}/R_0$, we find a definite trend in $T_0$ as Al% increases in the barrier: higher Al% leads to a lower $T_0$. It is also interesting to compare the trend of $T_0$ in the series of 20nm wide QWs in this study with a prior study on 10nm wide QW grown on (311)A GaAs orientation [21]: despite that the 10nm wide QW in [21] had 10% Al in the barrier, it showed $T_0$ higher than all the 20nm wide (001) samples here, reflecting the importance of having all other structural parameters consistent except changing only one parameter (Al%) in the present study. However, it is intriguing that the overall linear trend of $T_0$ vs. $p$ is similar and has almost the same slope between ref.[21] and this work..

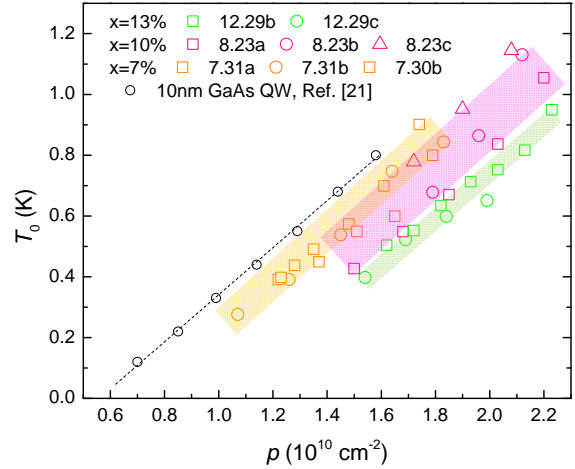

FIG 3. Characteristic temperature, $T_0$, vs 2D hole density for various GaAs QWs with different Al fraction in barrier, where $T_0$ represents the temperature at which $R_{xx}(T)$ peaks.

In Fig. 4 we also show that the overall shape of the temperature dependent electronic contribution to $R_{xx}$ is qualitatively unchanged throughout all our measurements, regardless of percentage of Al in the barrier. By fitting Eq.1 to our raw $R_{xx}$ data, we determined the residual resistance $R_0$ and the Bloch-Gruneisen coefficients. Using the fitted parameters, we determined the electronic contribution to $R_{xx}$ by subtracting $R_0$ and phonon contributions. To show a universality of the $R_{el}$ peak, this was completed for all Al%'s and various $p$'s. For an easier comparison without putting the data in the context of any specific model, we normalized the resistance and temperature with the height of the electronic contribution peak, $R_{peak} - R_0$, and $T_0$ respectively. We ascertain that the shape of the electronic contribution is independent of the microscopic details of the short-range disorder potential.

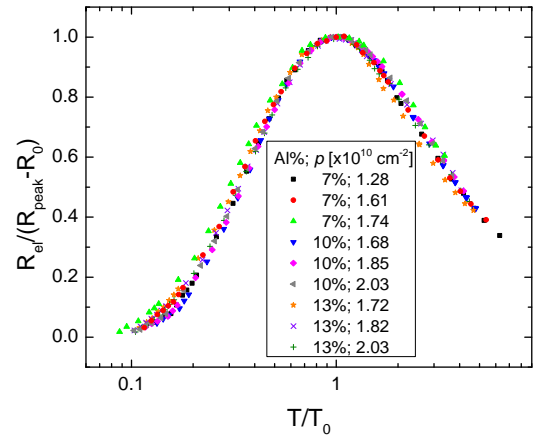

FIG.4. The dimensionless, $T$-dependent part of the electronic contribution to $R_{xx}$, normalized to the height of the resistance peak in $R_{xx}(T)$, plotted against temperature normalized with the characteristic temperature $T_0$.

We now discuss the implications of the experimental observations made here. In a previous work, Clarke *et al* pointed out that it is important to distinguish the short vs long range nature of dominant disorders in different semiconductor hetero-interface systems to resolve the conflicting metallicity strength [31]. Our present work explores the effect of systematically controlled short-range disorder strength from interface roughness and alloy scattering in *p*-GaAs, since all the samples have the same quantum well width and same or similar dopant set back distance except the Al% in the barrier. In addition, performing measurements to high temperatures ($T>T_F$), we show the universal shape of non-monotonic $R_{xx}(T)$ in samples exhibiting different strength of metallicity at low *T*. Up to date, the effect of short-range versus long-range (remotely ionized dopants) disorder on the non-monotonic $R_{xx}(T)$ from the high temperature ($T\sim T_F$) to the low temperature ($T\rightarrow 0$) regime has not been carefully addressed in relevant theories [7, 8, 27-29]. Given the consistent and significant impact of short-range scattering on both the quantitative strength ($R_{peak}/R_0$ differs by almost two times from $x=7\%$ to 13%) and characteristic temperature or energy scale of the 2D metallicity, it now becomes essential to differentiate the nature of disorder in quantitative theories about the metallic transport. Such effects could be readily incorporated and tested in Boltzmann transport based theories [27]. Within the Fermi liquid interaction correction theory [35] of the metallic transport, it is argued that stronger short-range disorder promotes large angle back scatterings of particles to interact with themselves and therefore the metallic correction should be stronger [31]. In a pioneering work by Ando [30], it was concluded that the interface roughness scattering strength is insensitive to Al% in GaAs/Al$_x$Ga$_{1-x}$As hetero-interface if a constant roughness parameter (typically a few Å) is assumed. Meanwhile, the short range alloy disorder potential in Al$_x$Ga$_{1-x}$As should be proportional to $x(1-x)$, i.e. an increasing function of *x* in our range of $x=0.07-0.13$. However, the alloy disorder scattering rate turns into a decreasing function of *x* after taking into account the exponentially suppressed carrier wavefunction within the Al$_x$Ga$_{1-x}$As barrier as *x* increases [30]. Therefore, one expects stronger short-range disorder scattering at smaller *x* values in our samples within the single particle relaxation model and our finding of stronger metallic resistivity drop at lower *x* agrees with Ref. [31]. Note that, the condition of $T_F \gg k_B T > \hbar/\tau$ required in the Fermi liquid interaction theory in the ballistic regime [35] of is not met in our experiment when the metallic conduction occurs [36] and the Hall coefficient in high mobility *p*-GaAs quantum wells shows anomalous behavior as compared to the Fermi liquid interaction theory [37]. Thus we do not attempt to further quantify the Fermi liquid parameter $F_0^\sigma$ by fitting data to theory.

In the literature, it is well understood that the mobility of high quality 2D carriers in modulation doped hetero-structures is limited by long-range disorder scatterings such as remote ionized impurities or ionized background impurities [32] but not the interface roughness or alloy scattering [30] in the low density regime. In our samples, we also observed a decreasing trend of mobility (in the $T\rightarrow 0$ limit) when the samples were gated to lower densities [34], indicating the dominance of long range Coulomb scattering as oppose to short-range disorder scattering in the low *T* mobility. It is thus striking to see that short-range disorder plays a clear role in setting the magnitude of the non-monotonic $R_{xx}(T)$ peak in dilute 2D carrier systems when the system is cooled down from $T\sim T_F$ to $T\ll T_F$.

The systematic shift of $T_0$ to higher scales in QWs with lower Al% fraction is intriguing. In all the theories invoking a non-monotonic $R_{xx}(T)$ in the metallic regime[7, 8, 27-29], this cross-over temperature $T_0$ is controlled by the Fermi temperature $T_F$ which is indeed linearly proportional to density *p* in 2D and comparable to the values of $T_0$ in experiment. However, our finding of $T_0 \propto p-p_c$ in Fig.3 with $p_c$ controlled by the disorder strength/type is more consistent with a two-component picture of the 2D metallic state involving a mobile component with density $p-p_c$ and more localized part with density $p_c$ [38]. In such perspective, $T_0$ should be controlled by the $T_F$ of the mobile component and thus naturally explains the appearance of the offset by $p_c$ in the linear proportionality between $T_0$ and *p*. Since our system resides in the low disorder regime (resistivity as low as $\sim (h/e^2)/50$), has strong correlations ($r_s\sim 20$), and the suppressed Hall response [37, 38], it is tempting to speculate that the localized component in the 2D metallic state arises from bubbles of a Wigner crystal in a Fermi liquid background [4, 7, 8]. How the viscosity of such a micro-emulsion phase of electron crystal and fluid depends on the disorder type will be an interesting theoretical issue [39] to compare with our experiments. We also point out that a recent two-component semi-classical effective medium theory for the 2D metallic state may also be consistent with this speculation [40]. A systematic and quantitative investigation of the crossover temperature $T_0$ and $R_{peak}/R_0$ versus short-range disorder strength and carrier density in such a theory would thus be desirable.

In summary, we have studied the impact of short-range disorder scattering on the non-monotonic temperature dependent resistance in strongly correlated 2DHS in 20nm wide GaAs QWs. By changing the Al fraction in the $Al_xGa_{1-x}As$ barrier, we were able to change the strength of the short-range disorder potential in strongly correlated ($r_s \sim 20$) 2D holes. The strength of the anomalous 2D metallic conduction was suppressed as the Al fraction was increased. Our analysis also shows that an increased Al fraction also leads to a suppressed characteristic temperature $T_0$ below which the metallic conduction occurs.

N.G. is partially supported by a US Department of Education GAANN fellowship (grant number P200A090276 and P200A070434). M.J.M. acknowledges support from the Miller Family Foundation. The molecular beam epitaxy growth at Purdue is supported by the U.S. Department of Energy, Office of Basic Energy Sciences, Division of Materials Sciences and Engineering under Award DE-SC0006671. X.P.A.G thanks the NSF for funding support (DMR-0906415) and Vlad Dobrosavljević for discussions. The authors also acknowledge a valuable insight on the short range disorder scattering strength of an anonymous referee.

———————

# Supplementary material

## Impact of Short-Range Scattering on the Metallic Transport of Strongly Correlated 2D Holes in GaAs Quantum Wells

Nicholas J. Goble,[1] John D. Watson,[2,3] Michael J. Manfra,[2,3,4,5] and Xuan P. A. Gao[1,*]

[1]*Department of Physics, Case Western Reserve University, Cleveland, Ohio 44106, USA*
[2]*Department of Physics, Purdue University, West Lafayette, Indiana 47907, USA*
[3]*Birck Nanotechnology Center, Purdue University, West Lafayette, Indiana 47907, USA*
[4]*School of Materials Engineering, Purdue University, West Lafayette, Indiana 47907, USA*
[5]*School of Electrical and Computer Engineering, Purdue University, West Lafayette, Indiana 47907, USA*

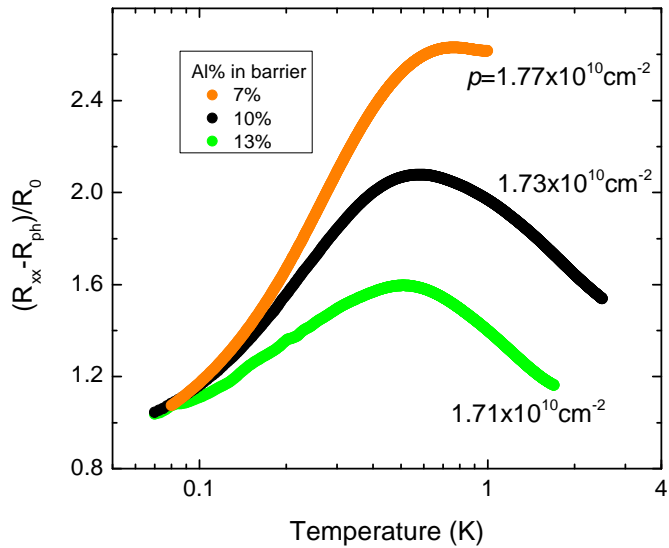

Fig.S1 Normalized resistance due to hole carrier scattering for different Al mole fractions in the $Al_xGa_{1-x}As$ barrier in 20nm wide GaAs/ $Al_xGa_{1-x}As$ quantum wells. Resistance from phonon scattering is determined using Eq. (1) in the main text and subtracted from $R_{xx}$.

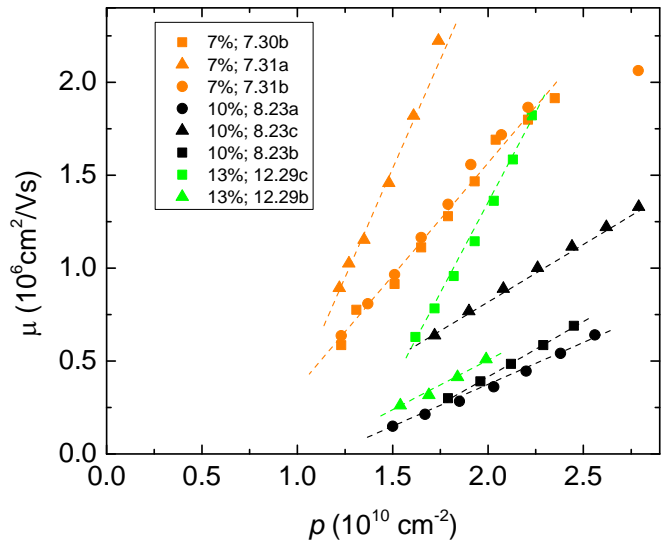

Fig.S2 Low temperature ($T\to 0$) mobility vs. hole density in 20nm wide GaAs/ $Al_xGa_{1-x}As$ quantum wells with different Al mole fractions in the $Al_xGa_{1-x}As$ barrier. Different symbol color corresponds to different $x$ in the $Al_xGa_{1-x}As$ barrier (orange: $x$=7%; black: $x$=10%; green: $x$=13%), and different symbol shape corresponds to different sample or current-voltage configuration. The dashed lines are a guide to the eye.